# DRSP : Dimension Reduction For Similarity Matching And Pruning Of Time Series Data Streams

Vishwanath R H[1], Samartha T V[1], Srikantaiah K C[1], Venugopal K R[1],
L M Patnaik[2]

[1]Department of Computer Science and Engineering, University Visvesvaraya College of Engineering, Bangalore University, Bangalore
[2]Honorary Professor, Indian Institute of Science, Bangalore, India

## ABSTRACT

*Similarity matching and join of time series data streams has gained a lot of relevance in today's world that has large streaming data. This process finds wide scale application in the areas of location tracking, sensor networks, object positioning and monitoring to name a few. However, as the size of the data stream increases, the cost involved to retain all the data in order to aid the process of similarity matching also increases. We develop a novel framework to addresses the following objectives. Firstly, Dimension reduction is performed in the preprocessing stage, where large stream data is segmented and reduced into a compact representation such that it retains all the crucial information by a technique called Multi-level Segment Means (MSM). This reduces the space complexity associated with the storage of large time-series data streams. Secondly, it incorporates effective Similarity Matching technique to analyze if the new data objects are symmetric to the existing data stream. And finally, the Pruning Technique that filters out the pseudo data object pairs and join only the relevant pairs. The computational cost for MSM is $O(l*ni)$ and the cost for pruning is $O(DRF*wsize*d)$, where DRF is the Dimension Reduction Factor. We have performed exhaustive experimental trials to show that the proposed framework is both efficient and competent in comparison with earlier works.*

## 1. INTRODUCTION

We live in a world driven by data. There are vast amounts of data being generated from the ubiquitous social networks with people constantly sharing pictures, videos and other multimedia content. Then there are legacy systems of banks and corporations that maintain all historical day-to-day transactions. We have Global Positioning Systems (GPS) with constant data interactions between the satellites and handheld devices to get location data and directions. There is data returned from sensor networks monitoring various parameters in the physical world. With vast amounts of data generation and flow in the current world, their understanding and analysis helps in gaining valuable insights into the relationships that exist between the data, the trend or the patterns represented by the data and thereby help to better comprehend and solve some real-world issues.

Data Mining is the process of gathering, comprehending and analyzing vast amounts of data to discover patterns and relationships. Time series data is a sequence of data points collected at regular intervals of time. Analysis of such time series data is called Time Series Analysis (TSA). This process of TSA helps in discerning the information present in the data and the knowledge gained by doing so can be used for similarity analysis and forecasting.

### A. Motivation

Stream Time Series data contain large set of time series data. A typical example is the data stream of the share prices of a company collected over multiple years. Analysis of such a data stream is vital for stock market analysts and investors to understand the demand, price and other such attributes to forecast the future trends. Another example is the time series data stream of the location of an object from a GPS device. Analysis of such data streams is vital in monitoring the movement of the object and detection of positions that are outliers from the normal trajectory of the object. However, the time series data streams are very large data sets with continuous inflow of data and time required for applications to perform analysis on them becomes high. Thus, there is a need to come up with an efficient methodology to reduce this large data stream into a relatively smaller data set, wherein all the vital statistics and pattern depicted by the initial large data stream is maintained in the compacted data set.

Moreover, there is also a need for a competent similarity matching technique which helps in depicting whether a new incoming data object is in-sync with the existing data stream or if it displays some outlier properties. This is significant as quite a number of decisions have to be made based on the result of similarity matching on the data object. For example, in case of the stock value analysis, any failure in the similarity matching marks the beginning of a different trend (falling or rising) in the stock value. The business analysts would use this inflection point to analyze the cause and result of such a change. In case of GPS data of the object, an outlier of similarity matching is an indication that the object is going beyond its normal trajectory or boundaries.

### B. Methodology

Dimension reduction can be achieved using Multi-level Segment Mean (MSM) [1]. This involves the process of fragmenting the data into equi-sized segments and obtaining a representative element for each segment of data. This process helps in reducing the number of data points in the data stream. Further, we use reduced data set for performing similarity matching using Sliding Window concept, which is a subset of the data stream and contains the most recent data elements. All comparisons for similarity matching are performed keeping the elements in this Sliding Window as the base parameters instead of all the elements of the entire data stream[2]. We update the Sliding Window with the new data object if it passes the similarity matching test. This reduces both the space and time complexity involved in the evaluation of similarity of the new incoming data object. The Sliding Window concept works well with moving data streams that have constant inflow of data at regular periods, like stock market data streams, GPS data streams, sensor data streams.

### C. Contribution

In this paper, we contributed a frame work called Dimension Reduction for Similarity Matching and Pruning of Stream Time Series data (DRSP) for addressing two major issues of (a) Dimension Reduction and (b) Similarity Matching and Pruning. Dimension Reduction is implemented using technique called Multi-level Segment Mean and Similarity Matching is achieved by applying Euclidean Distance equation. Pruning is performed on the matched pairs to eliminate pseudo or near-pseudo pairs (if any) present. Upon passing through the proposed framework, the best matched pairs of incoming data objects can be retrieved, with minimal space requirement and reduced time complexity.

### D. Organization

The paper is organized as follows - Section 2 discusses briefly the Literature on Dimension reduction and Similarity Search Techniques. Section 3 presents the Background work, Section 4 contains Problem Definition, Section 5 describes the System Architecture, Section 6 presents the Mathematical Model and Algorithm, Section 7 presents the Experimental Results. Concluding remarks are summarized in Section 8.

## 2. LITERATURE SURVEY

Similarity matching and searching is one of the greatest challenges in the area of mining time series data. This is due to the huge data size, number of dimensions and number of sequences that lead to a very expensive querying process. Many approaches have been proposed for the similarity search on archived time-series.

Agrawal et al., [3] developed whole sequence matching which finds sequences of the same length that are similar to a query sequence using Discrete Fourier Transform (DFT ). This approach works well only with sequences of the same length that are similar to a query sequence. Floutsos et al., [4] extended this work to subsequence matching using efficient indexing method to locate 1-dimensional subsequences within a collection of sequences. In both the works, Euclidean distance is used to measure the similarity among sequences. Tripti Negi et al. [5] proposed a model for identifying similar time series. A static index structure called kd-tree is used to index the time series and the Range search methodology is used to search for the similar subsequences. The indexing structure created using kd-tree is stored in the main memory. It is done considering the fact that the cost of the memory is decreasing rapidly. If the data is large and the index structure spills over to secondary memory the performance of the system suffers.

 Guttman et al., [6] designed a Dynamic Index structure for spatial searching. In order to perform an efficient search, the GEMINI framework is proposed to index time series and answer similarity queries without false dismissals. Since the dimensionality of time series are usually high (e.g., 1024), various Data reduction techniques have been proposed to reduce the dimensionality of time series before indexing them. Here only Discrete Fourier Transform(DFT ) and Discrete Wavelet Transform(DWT) have been used in the scenario of stream time series analysis. Xian et al., [7] introduced multiscale representations for fast pattern matching in stream time series using multiscale segment mean, which can be incrementally computed and perfectly adopted to the stream characteristics and also address the problem of matching both static and dynamic patterns over stream time-series data. This approach is not so suitable for image data streams. Sethukkarasi et al., [8] presented a technique for similarity matching between static/dynamic patterns and time-series image data to perform effective retrieval of image data.

Keogh et al., [9] have implemented Piecewise Aggregate Approximation, Adaptive Piecewise Constant Approximation, Symbolic Aggregate Approximation and a bit level time series representation techniques for pattern searching.

Li et al. [10] discussed similarity searching based on Piecewise Polynomial Representation i.e. to map each subsequence into a small set of multidimensional rectangles in feature space which is spanned by base of linear polynomial. Joshi [11][12] proposed a distinct similarity model for time series based on slope variations. In a preprocessing step, time series are normalized both on their duration and amplitude. After preprocessing, for small subsequences of equal length, the slopes are compared. For the similarity assessment, the cumulative variation in time-weighted slopes is computed. Their technique can handle vertical shifts, global scaling and shrinking as well as variable length queries. One shortcoming of the approach is the missing support of subsequence matching.

Temporal order and High Dimensionality are the two important characteristics of Time series data. Due to Temporal ordering of the time series, the consecutive values in the time series are related to each other, For example, for any Stock time series, the differences between consecutive values will be within some predictable threshold most of the time. This temporal relationship among nearby time series data points produces the redundancy, and hence such redundancy yields to data reduction. High Dimensionality in any Time series is considered as, each time point of a time series as a dimension, a time series is a point in a very high dimensions. A time series of any length n leads to a point in a n-Dimensional space. If we consider a time series of length n is 2000, which corresponds to a point in a 2000-dimensional space.

Dimension reduction is a crucial component for management of time series data. A number of dimension reduction techniques have been proposed for time series data sets. These techniques can be classified into two categories namely data adaptive and non data adaptive methods. In the data adaptive method, the dimensionality of the entire dataset is reduced and is typified by Principal Component Analysis PCA [13] and Singular Value Decomposition SVD [14]. PCA and SVD methods, generate optimal results but they involve high computational costs. PCA converts the raw data into a set of principal components with the first one having the largest variance, and each succeeding one having the highest variance under the constraint that it be orthogonal to the preceding components. Then the dimension reduction is performed by dropping last few components. In the case of SVD reduction of data is obtained by decomposing the data matrix.

Non data adaptive methods reduce the dimensionality of each data by a universal method which is not adaptive to concrete data such as Discrete Fourier Transform (DFT) [15, 4] and Data Reduction Technique (DRT) [16], Discrete Wavelet Transform (DWT) [4, 15], Random Mapping (RM)[18]. DFT and DRT transform raw data to frequency domain by expressing data in terms of sum of trigonometric functions and drop the low-frequency components to get the dimensionality of data reduced. DWT works similarly, and captures both frequency and location information by using a so-called wavelet function. RM just picks some dimensions randomly and simply maps raw data into this low-dimensional space.

Jeong et al., [19] present data reduction methods based on the DWT to handle large and complicated data curves. The methods minimize objective functions to balance the trade-off between Data reduction and modeling accuracy. Based on evaluation studies with popular testing curves and real-life datasets, these methods demonstrate their competitiveness with the existing

Data reduction and statistical data de-noising methods for achieving the Data reduction goals. Kim et al., [20] proposed the dimensionality reduction algorithm that can extract most salient and discriminative input features for output prediction. This approach is based on the Granger causality, a statistical technique, which aims at discovering a low-dimensional subspace that preserves the causality between input and output. Pritesh et al., [21] applied a decision tree technique in which each branch node represents a choice between alternatives and each node represent the decision or classification.

Nguyen et al., [22] present a dimensionality reduction method called IPIP is an Indexable Perceptually Important Points. This method takes full advantages of PIP (Perceptually Important Points) method with some improvements such that the new method can theoretically satisfy the lower bounding condition for time series Data reduction methods. Furthermore, they make IPIP indexable by showing that a time series reduced by IPIP can be indexed with the support of a multidimensional index structure based on Skyline index. IPIP method with its appropriate index structure can perform better than previous schemes, namely PAA based on traditional R*- tree. Seung-Hyun et al., [23] present a data reduction method for instance based learning, based on entropy-based partitioning and representative instances. This approach is not suitable for dimensionality or attribute reduction

Pruning is the process of eliminating the pseudo or unwanted data values from the data sources for the faster and accurate similarity search. There are several pruning techniques have been proposed by the researchers.

Chowdhury et al., [24] propose a tree-based candidate pruning technique HUC-Prune (high utility candidates prune) to efficiently mine high utility patterns without level-wise candidate generation-and-test. It exploits a pattern growth mining approach and needs maximum three database scans in contrast to several database scans of the existing algorithms.

## 3. BACKGROUND

Lian and Lei [2] proposed a method for Uncertain Stream Join (USJ) which addresses the problem of similarity matching and join on uncertain stream data. The authors propose a model to match the uncertain data streams, prune the results and finally join the best matched data pairs. To achieve this, the authors compare the new incoming data object with the entire data stream. It involves a high time and space costing process with large number of comparisons required to get the right matching pairs.

In this paper, we propose a frame work called Dimension Reduction for Similarity Matching and Pruning of Stream Time Series data (DRSP) to simplify the process of similarity matching and join of data streams by reducing the size of the original (base) data stream in an efficient way by employing Multi-level Segment mean (MSM) algorithm [1] such that the data pattern depicted by the original data stream is preserved. Furthermore, we adopt an effective similarity matching and pruning technique to filter out the pseudo pairs and retrieve only the nearest matching data pairs for joining. This reduces the space complexity and Computational cost associated with the storage and similarity matching of large time-series data streams.

# 4. PROBLEM DEFINITION

Consider two time series data streams $\overline{T_1}$ and $\overline{T_2}$ such that $\overline{T_1}$ = (R1, R2, R3, …, Rn) and $\overline{T_2}$ = (S1, S2, S3, …, Sn). This framework is applicable for multiple data streams as well. Consider the data streams to be periodic, i.e., the data objects contained in the data set correspond to regular periods in time and the two data streams are of equal length.

The first part of the proposed work aims at segmenting and obtaining a representative set of this large data stream. In order to recognize the symmetric pairs in the incoming data object, we do not have to consider the entire original data streams. It is sufficient if an effective representative set of the original data stream is generated. This is because, many a times, each and every data object in the data stream might not be significant and might contain sufficient of redundant data. Thus, a novel methodology to reduce the size of the data stream and the Similarity Matching and Pruning techniques are required. This data segmentation and reduction methodology is effective in retaining all the vital information present in the original data set. It produces a miniature version of the original data set by employing a methodology called MSM. Figure 1 shows the segmented and reduced data streams T1 = (r1, r2, r3, …, rt) and T2 = (s1, s2, s3, …, st). The data object rt+1 is the data object of T1 arriving at time instance t+1. Similarly st+1 is the data object of T2 arriving at time instance t+1. Then proceed to examine, if a new pair of data objects from the two time series data streams are symmetrical to the already present data objects in the two streams, i.e., to find out if the pair {rt+1 , st+1} are symmetrical to the data objects in T1 and T2 or if they are outliers. In case of outliers, then the framework helps to identify such pairs so that suitable action / warning can be raised in such cases. This reduced data set is used to perform Similarity Matching to get the data object pairs that are symmetrical to the existing data stream.

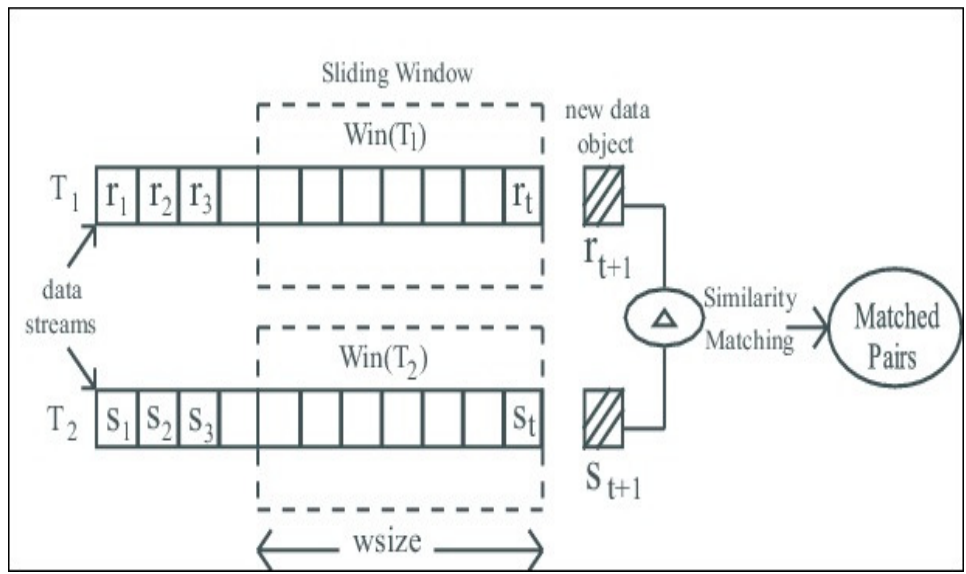

Figure 1 - Similarity Matching on the new data object

While comparing and matching the new data objects with the existing, pairs that match with the most recent of the data objects present in the data stream as against those that match with antiquated data objects are extracted. In other words, the objective is to retrieve pairs of new data objects which match the most recent trend present in the data stream. For this, a Sliding Window (*Win*) containing the most recent data objects of the data stream is used.  As shown in Fig 1, the Sliding Window contains the latest *window-size*(wsize) number of elements. At any time instance *t*, the Sliding Window of T$_1$ is represented as Win(T$_1$) = (r$_{t-wsize+1}$, r$_{t-wsize+2,}$ ..., r$_t$) where it contains the latest *wsize* data objects of  T$_1$. Similarly, the Sliding Window of T$_2$ at any time instance *t,* is represented as   Win(T$_2$) = (s$_{t-wsize+1}$, s$_{t-wsize+2,}$ ..., s$_t$) where it contains the latest *wsize*  data objects of  T$_2$. This Sliding Window is used as a representative data set for matching similarity of the new incoming data object. If the new data object pair {r$_{t+1}$ , s$_{t+1}$} is symmetric to the existing data objects in the Sliding Window, then this pair of data objects is appended to the corresponding data streams and the Sliding Window moves by one position such that         *Win*(T$_1$) = (r$_{t-wsize+2}$, r$_{t-wsize+3,}$ ..., r$_{t+1}$) and        *Win*(T$_2$) = (s$_{t-wsize+2}$, s$_{t-wsize+3,}$ ..., s$_{t+1}$) at time instance *t +1*. This process yields a representative set of the most recent data objects as the first element of the Sliding window is discarded and the new data object is included into the Sliding Window every time a new data object match is obtained, thereby keeping the Sliding Window updated with the most recent data objects of the data stream.

The Similarity Matching and Join [2] of data objects is formulated as given below.

***Similarity Matching and Join of data objects*** – Given two data streams T1  and T2 , a distance threshold  δ, Similarity Matching and Join of a pair of new data objects {ra , sb} is performed by comparing the new object pair with the elements in the Sliding Windows Win(T1) and Win(T2) such that dist{ra , sb} ≤ δ, where t-wsize+1 ≤ a,b ≤ t and wsize is the size of the Sliding Windows Win(T1) and Win(T2) which contain the latest elements of the data streams till time instance 't'.

The problem of Dimension reduction is solved via segmentation and similarity matching on a pair of new incoming objects and decide whether they are symmetrical to the existing objects in the data streams or not. Filter out the Matched pairs retrieved to get rid of the pseudo pairs present (if any) by pruning. The results (matched pairs) that are not eliminated at the end of the pruning technique are the ones that can be joined with the data stream. The data objects that are filtered by pruning technique are outliers that can be used for further analysis and detection of any discrepancy or anomaly.

## 5. SYSTEM ARCHITECTURE

The system architecture is depicted in Figure 2 and has the following components,
 (i) Input - Data Stream, (ii) Dimension reduction Engine, (iii)Similarity Matching and Pruning Engine, (iv)Output – Matched data object pairs

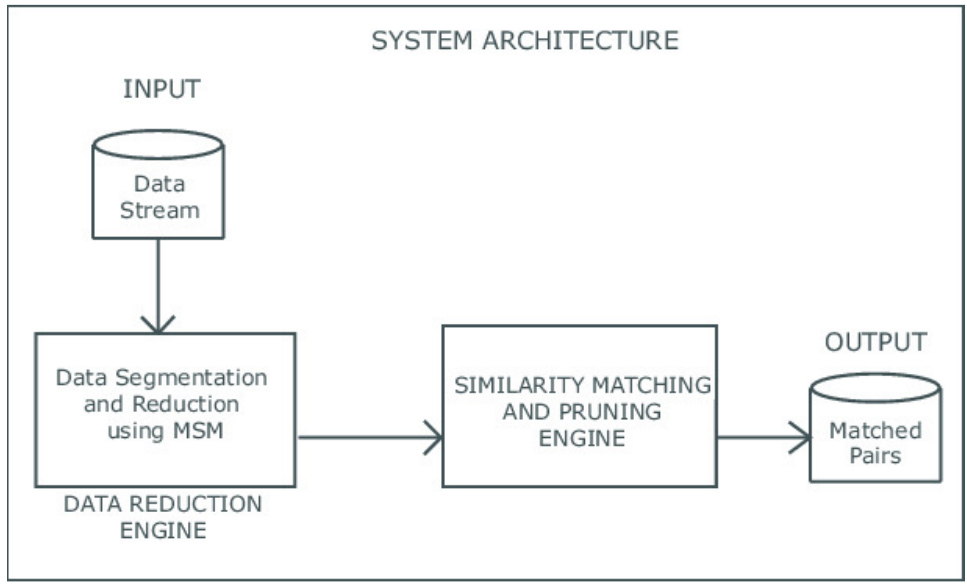

Figure 2 - System Architecture Diagram

***Input Data Stream –*** The input consists of time-series data streams. This might be time-series data streams of GPS locations of an object, data streams from sensor networks, Stock market data stream and so on.

***Dimension reduction Engine –*** The Dimension reduction Engine mainly performs the task of reducing the quantity of data while maintaining all the valuable information. This Engine uses a novel method called Multi-level Segment Means *(MSM)* to segment and reduce the data.

***Similarity Matching and Pruning Engine –*** This engine performs the task of matching similarity between the pairs of data objects from two data streams by using the concept of Euclidean Distance. It also further filters these data pairs to remove any pseudo pairs and retains only the symmetric pairs.

# 6. MATHEMATICAL MODEL AND ALGORITHM

### A. Data Segmentation and Reduction using Multi-level Segment Mean (MSM) Algorithm -

Data Reduction and Segmentation is the process of reducing the amount of records present in the data set. For example, consider a sensor to track the location of an object. This sensor sends location data of the object every 30 seconds. This constitutes to 2880 records per day. Most of these records might contain data that is redundant. It is not necessary have to analyze all the 2880 records to get the trajectory of the object for a particular day. The trajectory can be obtained by analyzing the important points in this data set. The Data reduction technique called Multi-level Segment Mean (MSM) is as shown in Figure 3. This technique helps to reduce the number of records in the process while retaining all the prime information and the pattern of the data flow as in the original data set. This technique is a consolidation of the work that has been discussed in [1].

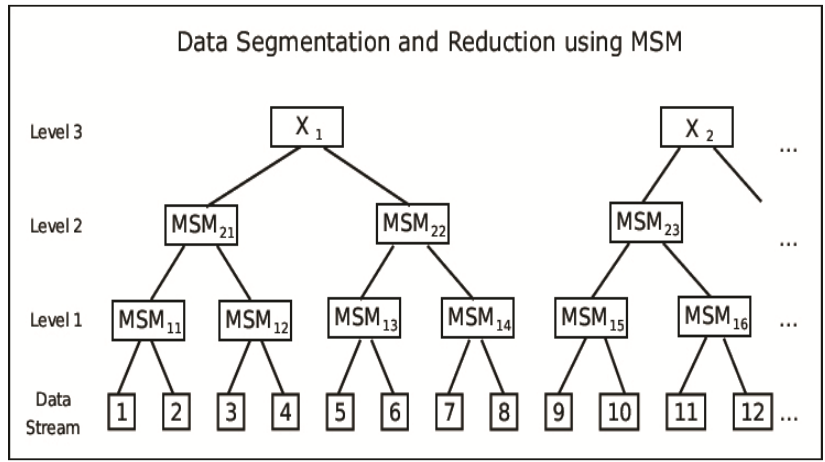

Figure 3 - Illustration of MSM

Consider a time series data stream, $T = (r_1, r_2, r_3, \ldots, r_i)$ and *length ( T ) = i*. In MSM, we split this time-series data set into equi-sized disjoint segments ($S_1$, $S_2$, …, $S_n$ ). The size of each segment is given by seg_size. Thus, the number of segments $n = i / seg\_size$. Then the mean of each segment is computed. This accounts to data approximation at level 1.

If the original data is represented by the set T after data approximation at level 1 as MSM1 , then
$$MSM_1 = ( MSM_{11}, MSM_{12}, \ldots, MSM_{1n}) \qquad \ldots..(1)$$

Any element of $MSM_1$ can be represented as given below

$$MSM_{1n} = \sum_{j=seg\_size(n-1)+1}^{seg\_size(n)} r_j / seg\_size \qquad \ldots..(2)$$

for all $n \in \{1, 2, \ldots, ( \text{length}(T) / seg\_size \}$ Eq(2) can be simplified as

$$MSM_{1n} = \frac{1}{seg\_size} * \sum_{j=seg\_size(n-1)+1}^{seg\_size(n)} r_j \qquad \ldots.(3)$$

Similarly $MSM_2$ can be represented as
$$MSM_2 = ( MSM_{21}, MSM_{22}, \ldots, MSM_{2n}) \qquad \ldots.. (4)$$

and expressed as follows

$$MSM_{2k} = \frac{1}{seg\_size} * \sum_{j=seg\_size(k-1)+1}^{seg\_size(k)} MSM_{1j} \qquad \ldots.. (5)$$

for all $k \in \{1, 2, \ldots, ( \text{length}(MSM_1) / seg\_size \}$

In general, the elements at any level *l* can be represented as

$$MSM_{lk} = \frac{1}{seg\_size} * \sum_{j=seg\_size(k-1)+1}^{seg\_size(k)} MSM_{(l-1)j} \qquad \ldots.. (6)$$

where $k = \{ 1, 2, \ldots, ( \text{length}(MSM_{l-1}) \,/\, seg\_size ) \}$ and          $MSM_0 = T$

The Input data set after applying MSM technique is reduced to $m$ records and is represented as

$$T = ( x_1, x_2, x_3, \ldots, x_m )$$          ……(7)

where  $m = length\ (\ T\ ) \,/\, seg\_size^l$          …… (8)

The time complexity for the MSM reduction process is $O(l*n_i)$ where $l$ is the number of levels of MSM and $n_i$ is the number of segments at each stage. The time complexity is directly proportional to the number of MSM levels. The number of levels upto which MSM can be applied needs to be chosen in such a way that the data pattern of the original data set is retained. If MSM is applied beyond a certain level (to compact the data to large extent), it fails to capture the pattern depicted by the original data set and Similarity Matching process will not yield desired results.

With MSM process, the data set would be compacted to a ratio of $1 : seg\_size^l$. This helps to reduce the space complexity of the data set by a factor of ( $1 \,/\, seg\_size^l$ ). This reduction in the data set further helps to reduce the time required for the computation of similar pairs.

Figure 4 is a plot of the hourly Stock values quotes of a particular company collected over 2 months. This constitutes to about 1440 data points.

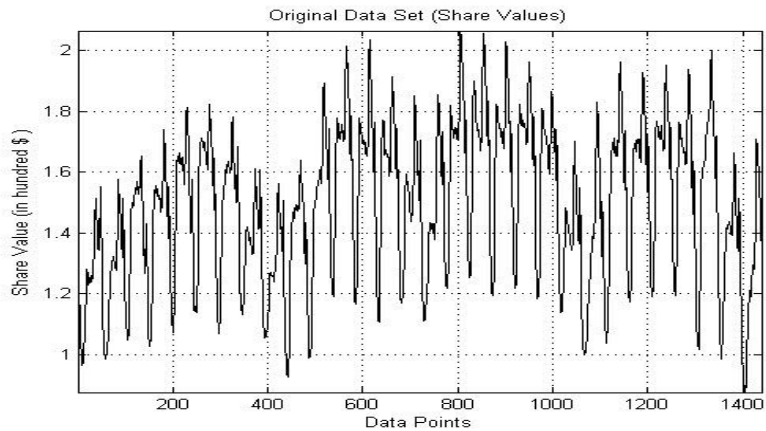

Figure 4 - Original data set – Share values

Figure 5 is a plot the reduced data set after applying MSM technique. In this example, segment size $seg\_size = 4$ and Number of levels $l = 2$. The original data set is compacted to a ratio of $1 : seg\_size^l$ i.e. $1:4$. Therefore, the reduced data set size is $1/4^{th}$ that of the original data-set.

From the plot in Fig 5, it can also be noted that the compacted data-set maintains all the major peaks and dips that are present in the original data set even though it has only 360 data points (records).

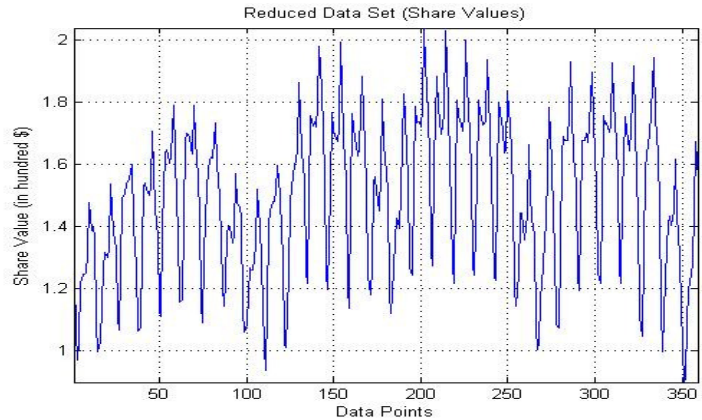

Figure 5 - Reduced data set – Share values

Thus, even after reducing the original data to a ratio of 1:4, the new data set retains the same pattern of data distribution as in the original data stream. The same MSM technique can be used to compress to higher ratios by changing the values of Segment size and Number of levels.

## B. Similarity Matching and Pruning

The task of matching similarity pattern between a pair of data objects is performed by employing the Euclidean Distance method. For any two time series data streams $T_1 = ( x_1, x_2, x_3, …, x_m)$ and $T_2 = ( y_1, y_2, y_3, …, y_m)$ the Euclidean Distance between the two is defined as the Cartesian distance between two data points in Euclidean space.

The process of Pruning performs the task of further filtering these data object pairs obtained after Similarity Matching and thereby retrieving only the nearest symmetric pairs while eliminating the pseudo pairs. In order to apply the Pruning Techniques, Sliding Windows $Win(T_1)$ and $Win(T_2)$ are constructed to obtain a set of most recent data objects of the two data streams $T_1$ and $T_2$ respectively. Now, consider a data pair $\{T_1(x_{t+1}), T_2(y_k)\}$ where $T_1(x_{t+1})$ is a new incoming data object at time stamp *t+1* beyond the Sliding Window of $T_1$ and $T_2(y_k)$ | ( *t* - $Win(T_2)$ + 2 ≤ k ≤ *t+1* )

which includes the data objects in the Sliding Window of $T_2$ as well as a new data object of $T_2$ at *t+1*. Then, checks are performed for the symmetry of a pair of data objects for $T_1$ and $T_2$ trimming the pairs that do not satisfy the Pruning Lemma, [2].

## Pruning Lemma -

*Given a pair of data Objects* $\{T_1(x_{t+1}), T_2(y_k)\}$ *and a Distance Threshold* δ, *the pair* $\{T_1(x_{t+1}), T_2(y_k)\}$ *can be pruned if* $Dist ( Cen(T_1(x_{t+1})), Cen(T_2(y_k)) ) - Rad(T_1(x_{t+1})) - Rad(T_2(y_k)) > δ$

For the application of the Pruning Lemma, construct a Hyper sphere around the new data object $T_1(x_{t+1})$ and $T_2(y_k)$, calculating it's *Center* $[Cen(T_1(x_{t+1}))]$ and *Radius* $[Rad(T_1(x_{t+1}))]$. Then compute the distance between the centers subtracting the radius of each hypeDDRSPhere and check if this is within the distance threshold (δ). If the distance is greater than the threshold, then the two objects are considered too far apart and hence can be pruned. Similarly, check is applied for new data object $T_2(y_{t+1})$ and a data object $T_1(x_k)$ | ( t - $Win(T_1)$ + 2 ≤ k ≤ *t+1* ).

Pruning is applied to filter out the pseudo pairs (if any) returned after the process of Similarity Matching and only the most symmetric data object pairs are retained. The algorithm Dimension Reduction for Similarity Matching and Pruning (DRSP) of Stream Time Series Data is discussed in Table 1.

The Data Segmentation and Dimension Reduction Engine along with Similarity Matching and Pruning Engine together constitute DRSP framework.

Table1. Algorithm: Dimension reduction for similarity matching and pruning (DRSP)

**Algorithm: DRSP ($\overline{T_1}$ , $\overline{T_2}$ , seg_size , MSM_levels , δ)**

Inputs : $\overline{T_1}$ , $\overline{T_2}$ – Original Data Streams

seg_size – segment size for MSM
MSM_levels – number of levels for MSM
δ – distance threshold
Output : MPairs - Matched pairs

**Phase 1 : Data Segmentation and Dimension Reduction**
N = seg_size, k = MSM_levels

(a) Split $\overline{T_1}$ into equi-sized segments S = {$S_1$, $S_2$, …, $S_n$}

Such that length of each segment = N
(b) for each segment $S_n$ in S
$D_{kn} = \sum S_n / N$
append $D_{kn}$ to list $D_k$
decrement k
end for
(c) Repeat (a) and (b) on $D_k$ till k = 0
$T_1 = D_0$ // reduced data stream

Repeat the same to reduce $\overline{T_2}$ to obtain $T_2$

**Phase 2 : Similarity Matching and Pruning**
(a) Compute Euclidean Distance for $T_1$ [t+1] and $T_2$ [t+1]
MPairs = pairs {$T_1$ [t+1], $T_2$ [t+1]} within δ
(b) Construct Sliding Windows $Win(T_1)$ and $Win(T_2)$
(c) for each pair in MPairs
(d) for i in $Win(T_2$ [i]) and j in $Win(T_1$ [j])
apply Pruning Lemma on {$T_1$[t+1] , $T_2$[i]}
apply Pruning Lemma on {$T_1$[j] , $T_2$[t+1]}
if pair {$T_1$[t+1] , $T_2$[t+1]} is not pruned
retain the pair {$T_1$ [t+1] , $T_2$ [t+1]} in MPairs
end if
else discard pair {$T_1$ [t+1] , $T_2$ [t+1]}
end else
end for
end for

# 7. EXPERIMENTAL RESULTS

In this first example, Dimension reduction  Similarity Matching and Pruning (DRSP) framework is applied on Stock Market time-series data set. The efficiency of DRSP technique and the improvements in performance over Uncertain Stream time series Join processing USJ [2] has been analyzed and evaluated.

***Stock Market Data –*** The closing share prices and the quantity of shares traded have been collected for a company from Newyork Stock Exchange (NYSE). This data has been collected for every half-hour time frame over a Quarter (3 months). Consider the time-series data stream of closing share price as T1 and the data stream depicting the quantity of shares traded as T2. Apply the Dimension Reduction Technique to reduce  this huge data set to a relatively smaller data set. Figure 6 is a plot of the Original data set of share values collected for a quarter and has 4400 data points (records). Figure 7 is a plot of the same data set after reduction by MSM technique and has 550 data points (records).

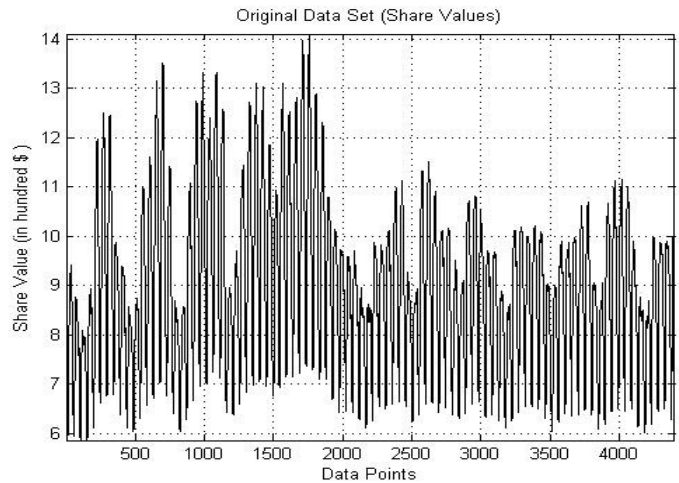

Figure 6 - Original data set – Share Values

Table 2. Comparison of percentage of matched pairs

| Type of Data | Data Set Size | % of Matched pairs |
|---|---|---|
| Original data set | 4400 | 85% |
| Reduced data set | 550 | 84% |

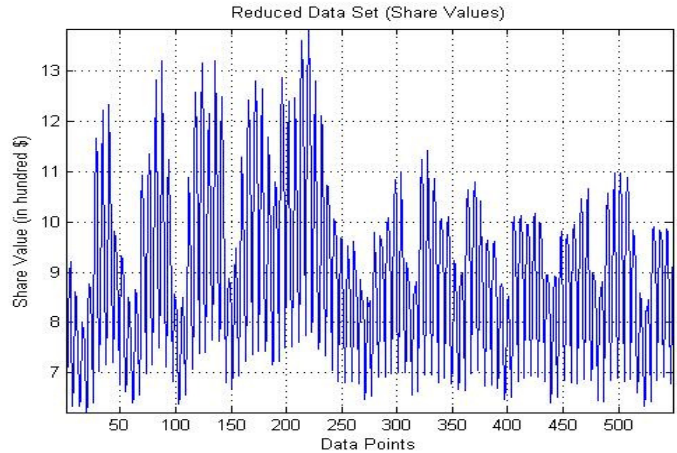

Fig 7 - Reduced Data set – Stock values

Table 2 represents the comparison between the number of matched pairs retrieved from the Original data set and the reduced data set. It shows that even after reducing the data to $(1/8)^{th}$ of the original data, the percentage of matched pairs is very close to that retrieved from the Original data set. It is observed that the Dimension reduction Technique effectively reduces the data while retaining most of the important information present in the data set.

By changing the MSM_levels and seg_size, the Dimension reduction Factor can be varied and thus the degree to which the data is reduction also varies. The Table3 below discusses the various percentages of Matched pairs retrieved for different degrees of reduction using MSM.

Table 3. Prediction of percentage of matched pairs for various segment size for MSM

| Original data Size | MSM level | seg_size | Reduced data size | % of Matched Pairs |
|---|---|---|---|---|
| 4400 | 3 | 2 | 550 | 85% |
| 4400 | 3 | 3 | 162 | 82% |
| 4400 | 3 | 4 | 69 | 80% |

Table 3 shows the Reduction of data size obtained by varying the seg_size for constant MSM_level. From the table, we can notice a trend wherein as the data size decreases, the percentage of matched pairs retrieved also reduces. It is observed that if the data is reduced to greater degrees, it fails to retain the pattern of the original data set.
 *Dimension Reduction Factor (DRF) = 1/seg_sizeMSM_level

Fig 8 is a plot of the time comparison for performing Pruning and obtaining the matching pairs for Original Data set and the Reduced Data set.

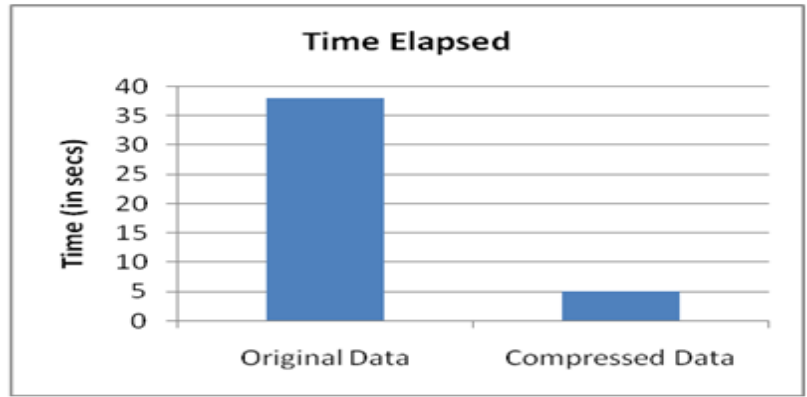

Figure 8 – Graph of time elapsed for Pruning

The time-complexity for Pruning ∝ $O(w.d)$ where $w$ is the window size of the Sliding Window and $d$ is the Data of the input data set. The Sliding Window is used to create a representative data set for comparison with the new incoming data object by using the concept of Euclidean Distance. The Sliding Window has the latest $win_{size}$ elements in it upon which the new data object is paired and checked for Object-level and Sample-level Pruning.

While executing Similarity Matching and Pruning on the Original Data Set, we set the $win_{size}$ to be 800. Given that we chose *MSM_levels = 3* and *seg_size = 2,* the *Data reduction Factor (DRF)* = (1/8). The same DRF is applied while constructing the Sliding Window on the Reduced Data Set as well. So, we set the $win_{size}$ for Reduced Data Set as 100 i.e. [(1/8)th of the $win_{size}$ of the Original Data Set].

DRSP framework was also applied to time-series GPS data, sensor data and synthetically generated data sets.

i) GPS Data to monitor the trajectory of an Object
ii)Sensor data to monitor the temperature and location of an object.
iii) Random Walk – A synthetically generated data set to monitor the movement of an object.

The GPS data set was collected over certain time duration to get the general trajectory of a vehicle between the source and destination. This trajectory helps define a pattern for the movement of the vehicle. This pattern would give information about the normal paths that the vehicle would take generally from the source to the destination. Any changes in the normal pattern of movement can be considered as outliers for the data set (signals for warning).The sensor data was collected from the heat sensors that would give the temperature of the object at regular time intervals. A GPS tracking device was also fixed to the object that would give the location of the object at regular intervals. This information from the sensor data stream and the GPS location stream can be used to monitor the movements of the object (like a robot) around a heat radiating source like a furnace. Outliers to this data set could be very high temperature readings from the sensor. This could be used to alert the operators and mark the boundaries for the movement of the object. And finally, we generated a synthetic data set for the random movement of an object in 2-d space and calculated its corresponding distance from a fixed point. If the object breached a particular perimeter, such data was considered as an outlier for the data set which could be used to generate warnings or alarms.

The reduced data set is obtained by applying MSM reduction technique with *MSM_levels = 3* and *seg_size = 2*. The *Dimension reduction Factor (DRF)* is $1/2^3 = 1/8$. Thus, the reduced data set is $(1/8)^{th}$ the size of the original data set. Thus, the MSM Dimension reduction technique we propose reduces the space complexity to DRF times that of the original data set.

Table 4. Comparison of percentage of matched pairs for different data sets

| Data Set | Type of Data Set | % of Matched pairs |
|---|---|---|
| GPS Data | Original Data Set | 80% |
| | Reduced Data Set | 79% |
| Sensor Data | Original Data Set | 88% |
| | Reduced Data Set | 88% |
| Random Walk | Original Data Set | 92% |
| | Reduced Data Set | 92% |

Table 4 shows the percentage of Matched pairs returned by applying Similarity Matching and pruning on Original data set and the Reduced data set by applying MSM technique. The Matched pairs returned by applying DRSP algorithm is almost same as that retrieved by applying Similarity matching and pruning on the Original Data Set. This further supports the fact that our Dimension reduction technique (MSM) maintains as many prime points from the original data set while reducing.

Table 5 shows the analysis performed by varying the Dimension Reduction Factor ( DRF ) for the data sets and the percentage of Matched Pairs retrieved from those data sets. Here the Original data is reduced by keeping the MSM_level = 3 and varying seg_size between 2 and 3. DRF = 1 / seg_sizeMSM_level.

Table 5. Comparison of percentage of matched pairs obtained for different reduction factors

| Data Set | Orig. data size | DRF | Reduced data size | % of Matched Pairs |
|---|---|---|---|---|
| GPS data | 4000 | 1/8 | 500 | 79% |
| | 4000 | 1/27 | 148 | 73% |
| Sensor data | 9000 | 1/8 | 1125 | 87% |
| | 9000 | 1/27 | 333 | 84% |
| Random Walk | 6000 | 1/8 | 750 | 92% |
| | 6000 | 1/27 | 222 | 90% |

The readings from Table 5 support the trend we noticed in our first experimental results wherein as the size of the Reduced data set reduces, the percentage of matched pairs retrieved also reduces. This is mainly due to the fact that as the size of Reduced data set becomes smaller, it fails to capture all the important information present in the Original data set. Thus, there is more loss of pattern while reducing to greater degrees and consequently, the application of Similarity matching and pruning on such a data set does not yield the best of the results. Thus it becomes imperative to choose the right value for DRF based on the kind of data values present in the Original data set.

The time complexity for Pruning $\propto O(\ w.d)$ where $w$ is the size of the Sliding Window and $d$ is the Data of the input data set. Sliding Window of the Reduced data set is $DRF$ times the size of Sliding Window of the Original data set. $win_{size}$ (reduced data set) = $DRF * win_{size}$ (original data set).

The calculation of Euclidean Distance for Similarity Matching and application of Pruning techniques all involve comparison of the new data Object with the members of the Sliding Window. With the reduction of the Window size, the number of key comparisons that the Similarity matching and Pruning Algorithms have to make to retrieve Matching Pairs from the data set reduces. Given that $win_{size}$ (reduced data set) $\propto DRF$, reduction in Window Size results in reduction in the time for computation of the Similarity Pairs and Pruning.

Also, while choosing the optimal $DRF$ value for reduction, it will be useful if the variance of the data values is also considered along with the size of the original data set. For example, if the variance of the data values in the original data set is minimal, then a slightly higher $seg\_size$ will not cause too much loss of data pattern while reduction. However, if the variance in the original data set is high, then increasing the $seg\_size$ would lead to greater loss of information in the reduced data set and similarity matching would not retrieve the right results.

Table 6 summarizes the commonly used symbols in this paper.

Table 6. Meanings of symbols used

| Symbol | Description |
|---|---|
| $\overline{T_1}$ , $\overline{T_2}$ | Input Stream Time Series Data sets |
| $T_1, T_2$ | Reduced data streams |
| *seg_size* | Segment size for MSM |
| *MSM_levels* | Number of levels for MSM |
| $\Lambda$ | Distance Threshold |
| MPairs | Number of matched pairs |
| $Win(T_1)$, $Win(T_1)$ | Sliding window of $T_1$ , $T_2$ Stream time series |
| $Cen(T_1)$, $Cen(T_2)$ | Centroid of the HypeDDRSPhere objects $T_1$ and $T_2$ |
| $Rad(T_1)$, $Rad(T2)$ | Radious of the HypeDDRSPhere objects $T_1$ and $T_2$ |
| *DRF* | Data reduction Factor |

# 8. CONCLUSION

In this paper, we have addressed the issue of reducing the dimension of stream and then applying Similarity Matching, Pruning and join process on the new incoming data object by proposing a framework called Data Segmentation and Similarity Matching (DRSP).

More specifically, we have proposed a novel methodology for Dimension reduction called Multi level Segment Mean (MSM) and have corroborated with experimental evidence about its effectiveness. MSM technique has a time complexity of $O(\ l * n_i\ )$ and reduces the Space Complexity of Pruning process to $DRF$ (Dimension Reduction Factor) times the Original. We have used varied time series data sets like GPS data, Stock market data, Sensor data and randomly generated synthetic data and have shown that the percentage of matched pairs retrieved from the compressed data set using MSM is very close to the percentage of matched pairs retrieved by applying similarity matching and pruning on the original data set. Our Framework reduces the time complexity of the Pruning technique to $O(DRF * w * d)$ where $DRF << 1$. The similarity matching technique can also be advanced such that it can filter out all the pseudo pairs which would obviate the need for a separate Pruning Engine.

## AUTHORS

Vishwanath R Hulipalled is an Assistant Professor at the Department of Computer Science and Engineering, Sambhram Institute of Technology, Bangalore, India. He received his Bachelor's degree in Computer Science and Engineering from the Karnataka University and Master of Engineering from UVCE, Bangalore University, Bangalore. He is presently pursuing his Ph.D in the area of Data Mining in JNTU Hyderabad. His research interest includes Time Series Mining, Financial Data Analysis, Sequence and Temporal Databases Analysis.

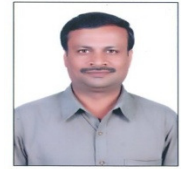

Samartha T V is currently a Research Assistant at the Department of Computer Science and Engineering, UVCE, Bangalore University, Bangalore. He received his Bachelor's degree in Computer Science and Engineering from Visvesvaraya Technological University, India. His research interests are Data Mining and Distributed Systems.

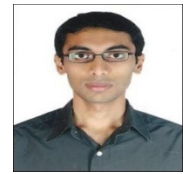

Srikantaiah K C is an Associate Professor in the Department of Computer Science and Engineering at S J B Institute of Technology, Bangalore, India. He obtained his B.E and M.E degrees in Computer Science and Engineering from Bangalore University, Bangalore. He is presently pursuing his Ph.D programme in the area of Web mining in Bangalore University. His research interest is in the area of Data Mining, Web Mining and Semantic Web.

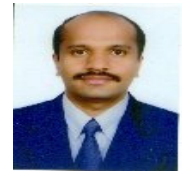

Venugopal K R is currently the Principal, University Visvesvaraya College of Engineering (UVCE), Bangalore University, Bangalore. He obtained his Bachelor of Engineering degree from UVCE. He received his Master's degree in Computer Science and Automation from Indian Institute of Science Bangalore. He was awarded Ph.D. in Economics from Bangalore University and Ph.D. in Computer Science from Indian Institute of Technology, Madras. He has a distinguished academic career and has degrees in Electronics, Economics, Law, Business Finance, Public Relations, Communications, Industrial Relations, Computer Science and Journalism. He has authored and edited 39 books on Computer Science and Economics, which include C Aptitude, Mastering C, Microprocessor Programming, Mastering C++, Digital Circuits and Systems, Petrodollar and the World Economy etc. During his three decades of service at UVCE, he has over 400 research papers to his credit. His research interests include Computer Networks, Wireless Sensor Networks, Parallel and  Distributed Systems, Digital Signal Processing and Data Mining.

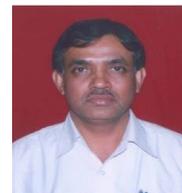

L M Patnaik is currently an Honorary Professor at Indian Institute of Science, Bangalore, India. He was formerly Vice Chancellor, Defense Institute of Advanced Technology, Pune, India. With more than three decades of service at the Institute, he has over 700 research publications in refereed International Journals and refereed International Conference Proceedings. He is a Fellow of all the four leading Science and Engineering Academies in India and Fellow of the IEEE and the Academy of Science for the Developing World. He has received twenty national and international awards; notable 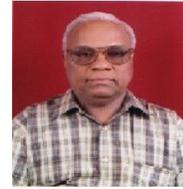 among them is the IEEE Technical Achievement Award for his significant contributions to High Performance Computing and Soft Computing. His research interest has been in the areas of Parallel and Distributed Computing, Mobile Computing, CAD for VLSI circuits, Soft Computing and Computational Neuroscience.